\documentclass[referee]{cjaa}           

\usepackage{graphicx}                   
\input{epsf.sty}                        
\input{psfig.sty}                       

\begin{document}

   \title{Relationship between the rise width and the full width of gamma-ray
burst pulses and its implications in terms of the fireball model
}

   \volnopage{Vol.0 (200x) No.0, 000--000}    
   \setcounter{page}{1}          

   \author{Rui-Jing Lu
      \inst{1,2,3}\mailto{}
\and Yi-Ping Qin \inst{1,2}
   \and Ting-Feng Yi
      \inst{2}
      }
   \offprints{R.-J. Lu et al.}                   

   \institute{National Astronomical Observatories/Yunnan
Observatory, Chinese Academy of Sciences, P. O. Box 110, Kunming,
Yunnan, 650011, P. R. China\\
\email{luruijing@126.com; ypqin@ynao.ac.cn}
        \and
          Physics Department, Guangxi University, Nanning, Guangxi 530004, P. R. China\\
        \and
         The Graduate School of the Chinese Academy of Sciences\\
          }

\date{Received~~2005 month day; accepted~~2005~~month day}
\abstract{Kocevski et al. (2003) found that there is a linear
relation between the rise width and the full width of gamma-ray
burst pulses detected by the BATSE instrument based on their
empirical functions. Motivated by this, we investigate the
relationship based on Qin et al. (2004) model. Theoretical analysis
shows that each of the two quantities, the rise width and the full
width of observed pulse, is proportional to
$\Gamma^{-2}\Delta\tau_{\theta,\rm FWHM}\frac{R_c}{c}$, where
$\Gamma$ is the Lorentz factor for the bulk motion, $\Delta \tau
_{\theta,\rm FWHM}$ is a local pulse's width, $R_{c}$ is the radius
of fireballs and c is the velocity of light. We employ the observed
pulses coming from four samples to study the relationship and find
that: (1) Merely the curvature effect could produce the relationship
with the same slope as those derived from Qin et al. (2004) model in
the rise width vs. the full width panel. (2) Gamma-ray burst pulses,
long or short ones (selected from the short and long GRBs), follow
the same sequence in the rise width vs. the full width panel, with
the shorter pulses at the end of this sequence. (3) All GRBs may
intrinsically result from local Gaussian pulses. These features
place constraints on the physical mechanism(s) for producing long
and short GRBs.  }

   \authorrunning{R.-J. Lu, \& Y.-P. Qin \& T.-F. Yi  }            
   \titlerunning{Relationship between the rise width and the full width and its implications }  

   \maketitle

%
%
\section{Introduction}           
\label{sect:intro}
Although the mechanism underlying the gamma-ray bursts is still an
unsolved puzzle, it is generally accepted that the large energies
and the short timescales involved require the gamma-rays to be
produced in a stage of fireball which expand relativistically
(see, e.g., Goodman 1986; paczynski 1986). An individual shock
episode gives rise to a pulse in the gamma-ray light curve, and
superposition of many such pulses creates the observed diversity
and complexity of light curves (Fishman et al. 1994). Therefore,
the temporal characteristics of these pulses hold the key to the
understanding of the prompt radiation of gamma-ray bursts. It is
generally believed that some well-separated individual pulses
represent the fundamental constituent of GRB time profiles (light
curves) and appear as asymmetric pulses with a fast rise and an
exponential decay (FRED), and many pulses have FRED-like shapes.

What result in the observed light curves? According to Ryde \&
Petrosian (2002), the simplest scenario accounting for the
observed GRB pulses is to assume an impulsive heating of the
leptons and a subsequent cooling and emission. In this scenario,
the rising phase of the pulse, which is referred to as the dynamic
time, arises from the energizing of the shell, while the decay
phase reflects the cooling and its timescale. However, in general,
the cooling time for the relevant parameters is too short to
explain the pulse durations and the resulting cooling spectra are
not consistent with observation (Ghisellini et al. 2000). As shown
by Ryde \& Petrosian (2002), this problem could be solved when the
curvature effect of the expanding fireball surface is taken into
account.

The diversities of gamma-ray light curves in morphology could be
interpreted within the standard fireball model (Rees \&
$M\acute{e}sz\acute{a}ros$ 1992), and the observed FRED structure
was found to be interpreted by the curvature effect as the observed
plasma moves relativistically towards us and appears to be locally
isotropic (e.g., Fenimore et al. 1996, Ryde \& Petrosian 2002;
Kocevski et al. 2003, hereafter Paper I). Several investigations on
modeling pulse profiles have previously been made (e.g., Norris et
al. 1996; Lee et al. 2000a, 2000b; Ryde \& Svensson 2000; Ryde \&
Petrosian 2002; Borgonovo \& Ryde 2001; Paper I), they derived
several flexible functions to describe the profiles of individual
pulses based on empirical or semi-empirical relations. E.g., as
derived in detail in Paper I, a FRED pulse can be well described by
the equation (22) or (28). Using this model, they found that there
is a linear relationship between the full width at half-maximum,
often denoted FWHM, and the rise width of gamma-ray burst pulses
detected by the BATSE instrument (see Fig. 10 in paper I), and the
same result can be found in the gamma-ray burst pulses detected by
the anti-coincidence shield of the spectrometer (SPI) of INTEGRAL
(see Fig. 5a Ryde et al. 2003).

Qin (2002) has derived in detail the flux function based on the
model of highly symmetric expanding fireball, where the Doppler
effect of the expanding fireball surface is the key factor to be
concerned, and then with this formula, Qin (2003) studied how
emission and absorbtion lines are affect by the effect. Recently,
Qin et al. (2004) presented the formula in terms of count rates.
Based on this model, some relations, a power law relationship
between the observed pulse width and energy (Qin et al. 2005a), an
anti-correlation between the power law index and the local pulse
width (Jia et al. 2005), and correlations between spectral lags and
some physical parameters, such as Lorentz factor and the fireball
radius (Lu et al. 2005a), have emerged. At the same time, some
characteristics have been found, such as a reverse S-feature curve
in the decay phase (Qin et al. 2005b) and an inflexion from
concavity to convexity in the rising phase (Lu et al. 2005b) of the
light curve determined by equation (21) in Paper II, The Combination
of these knowledge is suggestive of a potential relationship between
the rise width and the full width of the observed pulse, which
motivates us to investigate the relationship found by Kocevski et
al. based on Qin model and explore its implications in terms of the
fireball model.

Although the origins of short GRBs and long GRBs are not yet clear,
it is generally suggested that short GRBs are likely to be produced
by the merger of compact objects while the core collapse of massive
stars is likely to give rise to long  GRBs (see Zhang \&
$M\acute{e}sz\acute{a}ros$ 2004; Piran 2005). Many properties, such
as luminosity, $<V/V_{max}>$, the angular distribution, the energy
dependence of the duration, and the hard-to-soft spectral evolution,
even pulses profile of short GRBs , are also similar to those of
long GRBs (e.g., Schmidt 2001; Ramirez-Ruiz \& Fenimore 2000; Lamb
et al. 2002; Ghirlanda et al. 2004; Cui et al. 2005), which indicate
that the GRBs appear to have the same emission mechanism and
possibly different progenitors for long and short bursts. Motivated
by this, we also investigate the temporal structure of narrow pulses
with durations shorter than 1 s (FWHM) from short GRBs and long
GRBs.

This paper is organized as follows. In section 2, we investigate the
temporal characteristics of light curves of GRBs based on Qin model.
In section 3, we examine the relationship between the rise width and
the full width of observational pulses and explore its possible
implication in terms of the fireball model. Discussion and
conclusions will be presented in the last section.


\section{The Theoretical analysis}
\label{sect:Obs}
As derived in detail in paper II, the expected count rate of the
fireball within frequency interval $[\nu_1,\nu_2]$ can be
calculated with
\begin{equation}
C(\tau )=\frac{2\pi R_{\rm c}^3\int_{\widetilde{\tau }_{\theta ,\rm \min }}^{%
\widetilde{\tau }_{\theta ,\rm \max }}\widetilde{I}(\tau _\theta
)(1+\beta \tau
_\theta )^2(1-\tau +\tau _\theta )d\tau _\theta \int_{\nu _1}^{\nu _2}\frac{%
g_{0,\nu }(\nu _{0,\theta })}\nu d\nu }{hcD^2\Gamma ^3(1-\beta
)^2(1+k\tau )^2}.
\end{equation}
In above formula, $\tau _\theta$ is a dimensionless relative local
time defined by $ \tau _\theta \equiv c(t_\theta -t_{\rm c})/R_{\rm
c}$, where $t_\theta$ is the emission time in the observer frame,
called local time, of photons emitted from the concerned
differential surface $ds_\theta $ of the fireball ($\theta$ is the
angle to the line of sight), $t_{\rm c}$ is a constant which could
be assigned to any values of $t_\theta$, and $R_{\rm c}$ is the
radius of the fireball measured at $t_\theta=t_{\rm c}$; Variable
$\tau$ is a dimensionless relative time defined by $ \tau \equiv
[c(t-t_{\rm c})-D+R_{\rm c}]/R_{\rm c}$, where $D$ is the distance
of the fireball to the observer, and $t$ is the observation time;
$\widetilde{I}(\tau _\theta )$ represents the development of the
intensity magnitude in the observer frame, called as a local pulse
function; $g_{0,\nu }(\nu _{0,\theta })$ describes the rest frame
radiation mechanisms; And $k\equiv\beta/(1-\beta)$. At the same
time, the integral limits $\widetilde{\tau }_{\theta ,\rm \min }$
and $\widetilde{\tau }_{\theta ,\rm \max }$ are determined by $
\widetilde{\tau }_{\theta ,\rm \min }=\max \{\tau _{\theta ,\rm \min
},(\tau -1+\cos \theta _{\rm \max })/(1-\beta \cos \theta _{\rm \max
})\} $ and $ \widetilde{\tau }_{\theta ,\rm \max }=\min \{\tau
_{\theta ,\rm \max },(\tau -1+\cos \theta _{\rm \min })/(1-\beta
\cos \theta _{\rm \min })\},$ where $\tau_{\theta,\rm min}$ and
$\tau_{\theta, max}$ are the lower and upper limits of
$\tau_{\theta}$ confining $\widetilde{I}(\tau _\theta )$, and
$\theta_{\rm min}$ and $\theta_{\rm max}$ are determined by the
concerned area of the fireball surface, and then the radiation is
observable within the range of $(1-\cos \theta _{\rm \min
})+(1-\beta \cos \theta _{\rm \min })\tau _{\theta ,\rm \min } \leq
\tau \leq (1-\cos \theta _{\rm \max })+(1-\beta \cos \theta _{\rm
\max })\tau _{\theta ,\rm \max }$.

Formula (1) suggests that, light curves of sources depend mainly
on $\Gamma$, $\widetilde{I}(\tau _\theta )$ and $g_{0,\nu }(\nu
_{0,\theta })$. Observation suggests that the common radiation
form of GRBs is the so-called Band spectrum function (Band et al.
1993) which was frequently, and rather successfully, employed to
fit the spectra of the sources (see, e.g., Schaefer et al. 1994;
Ford et al. 1995; Preece et al. 1998, 2000), therefore we take in
this paper the Band function as the rest frame radiation form. And
the rise width and the full width of light curves depend on the
two factors of sources, $\Gamma$ and $\widetilde{I}(\tau _\theta
)$.

In this paper, we use $\tau_r$ and $\tau_{\rm FWHM}$ refer to the
rise width and the full width of light curves corresponding to
variable $\tau$, and $t_r$ and $t_{\rm FWHM}$ to those corresponding
to variable t, respectively. In the same way, we use $\Delta \tau
_{\theta,\rm FWHM}$ refer to the FWHM of a local pulse corresponding
to variable $\tau_\theta$, and $\Delta t _{\theta ,\rm FWHM}$ to
that corresponding to variable $t_\theta$.

For the sake of simplicity, we first employ a local pulse with a
power law rise and a power law decay to study this issue, which is
written as
\begin{equation}
\widetilde{I}(\tau _\theta )=I_0\{
\begin{array}{c}
(\frac{\tau _\theta -\tau _{\theta ,\rm \min }}{\tau _{\theta
,0}-\tau _{\theta ,\rm \min }})^\mu \qquad \qquad \qquad \qquad
\qquad (\tau _{\theta ,\rm \min }\leq
\tau _\theta \leq \tau _{\theta ,0}) \\
(1-\frac{\tau _\theta -\tau _{\theta ,0}}{\tau _{\theta ,\rm \max
}-\tau _{\theta ,0}})^\mu \qquad \qquad \qquad \qquad (\tau _{\theta
,0}<\tau _\theta \leq \tau _{\theta ,\rm \max })
\end{array}.
\end{equation}

Where $\tau _{\theta ,0}$ and $\mu$ are constants. The FWHM of this
local pulse would be $\Delta \tau _{\theta ,\rm
FWHM}=(1-2^{(-1/\mu)})(\tau _{\theta ,\rm \max }-\tau _{\theta ,\rm
\min })$. The relationships between $\tau_r$, $\tau_{\rm FWHM}$ and
$\Gamma$, and that between $\tau_r$, $\tau_{\rm FWHM}$ and $\Delta
\tau _{\theta ,\rm FWHM}$ for the light curves determined by
equation (1) are plotted in Fig. 1, and presented in Fig. 2 is the
relationship between the $\tau_r$ and the $\tau_{\rm FWHM}$ of the
light curves.

Figure 1 shows that the $\tau_r$ and $\tau_{\rm FWHM}$ decrease with
the Lorentz factor following $ \tau_r \propto \Gamma^{-2}$ and $
\tau_{\rm FWHM} \propto \Gamma^{-2} $, and they increase with
$\Delta \tau _{\theta ,\rm FWHM}$ following $\tau_r \propto \Delta
\tau _{\theta ,\rm FWHM}$ for every value of $\Delta \tau _{\theta
,\rm FWHM}$, and $\tau_{\rm FWHM} \propto \Delta \tau _{\theta ,\rm
FWHM}$ when $\Delta \tau _{\theta ,\rm FWHM} \geq$ 1. Thus we obtain
\begin{equation}
\tau_r = k_{1}\Gamma^{-2}\Delta \tau _{\theta ,\rm FWHM}=k_{1}p,
\qquad \qquad \qquad \qquad \qquad \qquad \qquad
\end{equation}
\begin{equation}
\tau_{\rm FWHM} = k\Gamma^{-2}\Delta \tau _{\theta ,\rm FWHM}=kp
\qquad \qquad \qquad  (\Delta \tau _{\theta ,\rm FWHM} \geq 1),
\end{equation}
where $p=\Gamma^{-2}\Delta \tau _{\theta ,\rm FWHM}$.
$k_{1}=0.597\pm0.006$ and $k=1.335\pm0.036$ for this local pulse.
Study reveals that each of the two quantities, $\tau_r$ and
$\tau_{\rm FWHM}$, is proportional to p, but independent of $\Gamma$
or $\tau _{\theta ,\rm FWHM}$.

Considering the relation between $\tau$ and t, we get from (3) and
(4) that
\begin{equation}
t_r = k_{1}p\frac{R_{\rm c}}{c}, \qquad \qquad \qquad \qquad \qquad
\qquad \qquad
\end{equation}
\begin{equation}
t_{\rm FWHM} = kp\frac{R_{\rm c}}{c} \qquad \qquad \qquad  (\Delta
\tau _{\theta ,\rm FWHM} \geq 1).
\end{equation}
We find from the left panel in Fig. 2 that the $\tau_r$ increases
linearly with the $\tau_{\rm FWHM}$ when we take $\Delta \tau
_{\theta ,\rm FWHM}$ = constant and $\Gamma$ = variable. We thus
perform a linear least square fit to the two quantities, and have
log($\tau_r$) = A + Blog($\tau_{\rm FWHM}$) for a certain value of
$\Delta \tau _{\theta ,\rm FWHM}$. There are almost the same slopes
of ``B" (i.e. B = 1.0) for different values of $\Delta \tau _{\theta
,\rm FWHM}$, whereas the value of ``A" changes from -1.981 to -0.526
when $\Delta \tau _{\theta ,\rm FWHM}$ changes correspondingly from
0.001 to 0.1, and there is a upper limit value of A$\simeq$-0.40
when $\Delta \tau _{\theta ,\rm FWHM}\geq1$, thus a dead line can be
found when $\Delta \tau _{\theta ,\rm FWHM}\geq1$, i.e.,
log($\tau_r$) = -0.404 + 1.00$\times$log($\tau_{\rm FWHM}$). (for
all situations of fitting, their correlation coefficient R $>$0.999
and the number of the data N=27.)

According to equations (6) and (7) in Paper II, one could find that
the photons that observer receives at different observation time
$\tau$ emit from the different surface of fireball when $ \Delta
\tau _{\theta ,\rm FWHM} < 1$, so that the profiles of the light
curves determined by formulas (1) are affected by $\Delta \tau
_{\theta ,\rm FWHM}$. Whereas when $\Delta \tau _{\theta ,\rm
FWHM}\geq1$, the photons reaching the observer at different
observation time $\tau$ come from the same whole surface of the
fireball, in this situation, the profiles of the light curves don't
change with $\Delta \tau _{\theta ,\rm FWHM}$. This analysis is
supported by the fact that the $\tau_r$ is sensitive to the $\Delta
\tau _{\theta ,\rm FWHM}$, but the $\tau_{\rm FWHM}$ isn't
significantly affected by the $\Delta \tau _{\theta ,\rm FWHM}$,
when $ \Delta \tau _{\theta ,\rm FWHM} < 1$, in fact in this case
the $\tau_{\rm FWHM}$ slightly decreases with the $ \Delta \tau
_{\theta ,\rm FWHM}$, and when $ \Delta \tau _{\theta ,\rm
FWHM}\rightarrow0$, the local pulse becomes a $\delta$ one, and the
$\tau_{\rm FWHM}$ would be determined by equation (44) in Paper II.
However when $\Delta \tau _{\theta ,\rm FWHM}\geq1$, each of the two
quantities, $\tau_r$ and $\tau_{\rm FWHM}$, linearly increases with
the $\Delta \tau _{\theta ,\rm FWHM}$ with the same slope (see the
right panel in Fig. 1). Which naturally explains why one could find
a dead line in the $\tau_r$ - $\tau_{\rm FWHM}$ panel when $\Delta
\tau _{\theta ,\rm FWHM}\geq1$.

When changing frequency interval from 100 $\leq$ $\nu/\nu_{0,p}$$
\leq$ 300 to 25 $\leq$ $\nu/\nu_{0,p}$$ \leq$ 50, or to other
frequency interval, and repeating the same work as above, we find
that the results don't change significantly, and the dead line is
not sensitive to frequency interval.

We study other forms of local pulses, such as $\mu=1, 3$ of equation
(2), an exponential rise and exponential decay pulse, a Gaussian
pulse, and a rectangle pulse, and so on, and find that the equation
(3) - (6) hold for all local pulses we investigate, and there are
different values of $k_{1}$ and k for different local pulse forms
(see Table 1). For every local pulse form, a dead line can be found
when $\Delta \tau_{\theta,FWHM}\geq1$. The dead lines of the six
local pulses, log($\tau_r$) = (-0.563 $\pm$ 0.005) +(1.015 $\pm$
0.002)log($\tau_{\rm FWHM}$) for local exponential pulse,
log($\tau_r$) = (-0.520 $\pm$ 0.009) +(1.027 $\pm$
0.003)log($\tau_{\rm FWHM}$) for local Gaussian pulse, log($\tau_r$)
= (-0.450 $\pm$ 0.003) +(1.005 $\pm$ 0.001)log($\tau_{\rm FWHM}$)
for $\mu=3$ of equation (2), log($\tau_r$) = (-0.404 $\pm$ 0.006)
+(1.007 $\pm$ 0.001)log($\tau_{\rm FWHM}$) for $\mu=2$ of equation
(2), log($\tau_r$) = (-0.318 $\pm$ 0.005) +(1.011 $\pm$
0.001)log($\tau_{\rm FWHM}$) for $\mu=1$ of equation (2), and
log($\tau_r$) = (-0.221 $\pm$ 0.011) +(1.027 $\pm$
0.003)log($\tau_{\rm FWHM}$) for local rectangle pulse, are
presented in the right panel in Fig. 2.

Studies show that, for all kinds of local pulse forms, the slopes of
their dead lines are always equal to 1.0, but there are different
intercepts for different local pulse forms in the $\tau_r$ -
$\tau_{\rm FWHM}$ panel. The intercept could therefore become an
indicator of the local pulse form. We also note that the dead line
of local rectangle form is the upper limit one for all local pulse
forms (i.e., the intercept of a dead line for any local pulse forms
would never exceed -0.20), which might be a criterion to check if
the temporal behaviors of gamma-ray burst pulses do result from the
contributions from the Doppler effect.

\section{Relationship between the rise time and the width of observational pulses}
\label{sect:data}
Kocevski et al. (2003) found that there is a linear correlation
between the rise width vs. the full width of gamma-ray burst pulses
provided by the BATSE instrument on board the CGRO (Compton Gamma
Ray Observatory) spacecraft. For the sake of convenience of
comparision with those obtained theoretically above, unlike Kocevski
et al., we here investigate the temporal structures of the light
curves of 2nd and 3rd channels based on their sample, respectively,
which we call sample 1. As they pointed out that a power-law rise
model can better describe the majority of the FRED pulses, so we
measure the $t_r$ and the $t_{\rm FWHM}$ of the pulses by fitted
with the equation of (22) in their paper. The results are presented
in Fig. 3.

Fig. 3 shows that the $t_r$ increase linearly with the $t_{\rm
FWHM}$ of the observed pulses. We perform a linear least square fit
at 1$\sigma$ confidence level to the two quantities, and have
log($t_r$)=(-0.531 $\pm$ 0.022) + (1.045 $\pm$ 0.029)log($t_{\rm
FWHM}$) with a linear correlation coefficient of 0.973 and a chance
probability of $p < 10^{-4}$ for the 2nd channel, and
log($t_r$)=(-0.492 $\pm$ 0.029) + (1.031 $\pm$ 0.024)log($t_{\rm
FWHM}$) with a linear correlation coefficient of 0.980 and a chance
probability of $p < 10^{-4}$ for the 3rd channel. The results show
that the two sequence in the $t_r - t_{\rm FWHM}$ panel have almost
the same intercepts and slopes within their error. Intriguingly, one
find that the slope is equal to the one obtained theoretically
above, which indicates that the observed results are well consistent
with those predicted by Qin model.

We thus may come to the conclusions: (1) Merely the curvature effect
can produce the relationship in the $t_r - t_{\rm FWHM}$ panel which
is independent of any frequency interval. (2) The gamma-ray burst
pulses most probably arise from local Gaussian form because the
sequences in the $t_r - t_{\rm FWHM}$ panel is closed to the dead
line of local Gaussian form. Note that the differences between the
two panels, $\tau_r$ - $\tau_{\rm FWHM}$ and $t_r - t_{\rm FWHM}$,
don't affect the comparison (See section (4)).

To further demonstrate these conclusion, we choose another sample,
call sample 2, based on gamma-ray burst pulses detected by HETE-2
instrument. Like Kocevski, we select pulses from the HETE-2 burst
home page (http://space.mit.edu/HETE/Bursts/) with the simple
criteria that pulses behave clean, well distinguished FRED-like
form, thus 12 pulses could be available in our sample 2. We measure
the $t_r$ and $t_{\rm FWHM}$ of these pulses with the same methods
as above. And the results are presented in Fig. 4. We perform a
linear least square fit to the two quantities with the same methods
adopted in Fig. 3, and obtain log($t_r$)=(-0.494 $\pm$ 0.065) +
(1.012 $\pm$ 0.087)log($t_{\rm FWHM}$) with a linear correlation
coefficient of 0.960 and a chance probability of $p < 10^{-4}$ for
the band B, and log($t_r$)=(-0.504 $\pm$ 0.059) + (1.088 $\pm$
0.078)log($t_{\rm FWHM}$) with a linear correlation coefficient of
0.976 and a chance probability of $p < 10^{-4}$ for the band C. The
results are well consistent with those obtained from Fig. 3, which
further testifies that the sequences in the $t_r-t_{\rm FWHM}$ plane
is independent of any frequency interval.

The first two smaller standard deviations of the four intercepts in
the two sample from the ones of the six dead lines obtained above
theoretically are 0.043 for local Gaussian pulse and 0.120 for local
exponential pulse, which indicates that these four sequences are
very closed to the dead line of the local Gaussian pulse and
implicate that these observed pulses may arise from local Gaussian
pulses. The conclusion is only a preliminary one which needs to be
confirmed by larger samples in the future.

As shown in Fig. 3 and 4, all pulses, which are selected from long
bursts, are longer than 0.5 s. Norris et al. 2001 pointed out that
short bursts with $T_{90}<2.6s$ have different temporal behaviors
compared with long bursts. Whether or not short pulses and long ones
follow the same sequence in the $t_r-t_{\rm FWHM}$ plane and behave
the same temporal behavior? motivated by this, we select 6 short
pulses, call sample 3, from the 64 ms count data of 532 short GRBs,
and 23 short pulses (shorter than 1 s), call sample 4, from long
GRBs with the same criteria adopted in sample 2. These short and
long GRBs are detected by the BATSE instrument. We repeat the same
work as above, and the results are plotted in the left panel in Fig.
5.


We find from the left panel in Fig. 5 that, all pulses (selected
from short and long GRBs) follow the same sequence in the
$t_r-t_{\rm FWHM}$ plane, with the shorter pulses at the end of this
sequence, which show that short pulses (or bursts) behave the same
temporal behaviors as long pulses, the only difference is that the
quantity of p (see equation of (3) or (4)) of short pulse is smaller
than the one of long pulse.

\section{discussion and conclusions }
\label{sect:data}

All analyses in this paper are based on formula (1), which is based
on the case of a fireball expanding isotropically with a constant
Lorentz factor, $\Gamma>1$. In fact, the symmetry of expansion
matters only over angles the order of a few times $\Gamma^{-1}$, and
beaming prevents us from observing other regions of the shell. The
formula (1) is suitable for describing light curves of spherical
fireballs or uniform jets. When considering a uniform jet and taking
$\theta_{\rm max}=\Gamma^{-1}$ in the formula (1), we measure the
$\tau_r$ and $\tau_{\rm FWHM}$ of the resulting light curves. They
show no difference from those of the spherical geometry, which
indicate that the relationships in the $\tau_r$ - $\tau_{\rm FWHM}$
plane are independent of gamma-ray burst pulses coming from either
spherical fireballs or uniform jets.

It is an ubiquitous trend that indexes of spectra of many GRBs are
observed to vary with time (see Preece et al. 2000). We wonder how
the relationship would be if the rest frame spectrum develops with
time. As Preece et al. (2000) pointed out that the typical fitted
value distributions for the low energy spectral index ($\alpha$) is
-1.5 $\sim$ -0.3, the high energy spectral index $\beta$ is -2
$\sim$ -3. Therefore here some sets of typical values of the indexes
such as (-1.5, -2), (-0.3, -3) and (-0.8, -2.5) would be employed to
investigate the relationship. Calculations show that the two
quantities, $\tau_r$ and $\tau_{\rm FWHM}$, are not significantly
affected by the rest frame radiation form.

We find that, owing to the Doppler effect of the fireball surface
(or the curvature effect), for any local pulse form, the width of
the light curve would always be $t_{\rm
FWHM}=k\Gamma^{-2}\Delta\tau_{\theta,\rm FWHM}\frac{R_{\rm c}}{c}$
(There are different values of k for different local pulse forms,
see Table 1). As derived in detail in Paper II, the width of the
light curve of the local $\delta$ function pulse would be $\tau_{\rm
FWHM}\simeq\frac{\sqrt{2}-1}{2}\Gamma^{-2}\Delta\tau$ (see (48) in
Paper II), where $\Delta\tau$ is the interval of the observable time
of the local $\delta$ function pulse, i.e.,
$\Delta\tau=1+\beta\tau_{\theta,0}$. When considering the
relationship between $\tau$ and t, and taking $\tau_{\theta,0}=0$,
we obtain $t_{\rm
FWHM}\simeq\frac{\sqrt{2}-1}{2}\Gamma^{-2}\frac{R_{\rm c}}{c}\simeq2
s (\frac{R_{\rm c}}{10^{15}cm})(\frac{\Gamma}{10^{2}})^{-2}$, which
would be the lower limit of the width of light curves for any local
pulse form. Because of the relativistic beaming of the moving
radiating particles, only the emission from a narrow cone with an
opening angle of $\Gamma^{-1}$ is observed, Ryde \& Petrosian (2002)
obtained that the curvature timescale resulting from relativistic
effects is $\tau_{ang}=1.7 s (\frac{R_{\rm
c}}{10^{15}cm})(\frac{\Gamma}{10^{2}})^{-2}$ (See (5) in their
paper). Even as they pointed out that this is a lower bound for the
observed duration of a pulse.  Thus it can be seen that the
curvature timescale and the lower limit of the width of light curves
are comparable.

According to the equations (3) and (6), we know that the only
difference between the two panels, $t_r-t_{\rm FWHM}$ and
$\tau_r-\tau_{\rm FWHM}$, is that the different observed pulses
resulting from the same $\Gamma$ and $\Delta\tau_{\theta,\rm FWHM}$
but from different values of $R_{\rm c}$ would be only corresponding
to one point in the $\tau_r-\tau_{\rm FWHM}$ plane; but in the
$t_r-t_{\rm FWHM}$ plane these observed pulses would be
corresponding to different points as they come from different values
of $R_{\rm c}$, and these different points must be on a line with
the slope, B=1.0. However the intercepts of the light curves in the
two panels dependent only on both forms and widths of their
corresponding local pulses.

It is widely accepted that gamma-ray bursts arise from the internal
shocks at a distance of $R_{\rm c}\sim10^{13}-10^{17}cm$ in the
scenario of standard fireball model (see Rees \&
$M\acute{e}sz\acute{a}ros$ 1992, 1994; $M\acute{e}sz\acute{a}ros$ \&
Rees 1993, 1994; $M\acute{e}sz\acute{a}ros$ 1995; Katz 1994;
Paczynski \& Xu 1994; Sari \& Piran 1997; Piran 1999; Spada et al.
2000; Ryde \& Petrosian 2002; and Piran 2005). To compare theoretic
conclusions with the observed results in the $t_r-t_{\rm FWHM}$
plane, we merge the left panel in Fig. 5 into the right panel in
Fig. 2 by applying equation (5) and (6) and taking $R_{\rm
c}=3\times10^{15}cm$, and the results could be plotted in the right
panel in Fig. 5. The results are in good agreement with those
predicted by Qin model. We notice that the all observed pulses are
under the dead line of local Gaussian form within their errors.

As shown above, local pulses' widths, $\Delta\tau_{\theta,\rm
FWHM}\geq0.1$, for most of observed pulses. Applying the
relationship between $\tau_{\theta}$ and $t_{\theta}$, we obtain
that $\Delta t_{\theta,\rm FWHM}$$\geq0.1\frac{R_{\rm c}}{c}$.
Because taking $\tau_{\theta,\rm min}=0$ (i.e., $t_{\theta,\rm
min}=t_{\rm c}$) in above analysis, we know that $R_{\rm c}$ at that
time is the radius of fireball at which radiation begins to emit,
which may correspond to the stage the fireball become optically thin
and photons can escape freely. And the fact that
$\Delta\tau_{\theta,\rm FWHM}\geq0.1$ indicates that local pulse's
width is not less than 1 order of the time scale the fireball
expands to become optical thin for most of observed pulses, and its
implication is not clear now because we don't know what result in
the local pulses yet.

In all, we come to the following conclusions: (1)Merely the
curvature effect could produce the observed relationship between
$t_r$ and $t_{\rm FWHM}$. (2)Both long pulses and short ones follow
the same sequence in the $t_r-t_{\rm FWHM}$ plane. If all observed
pulses come from the same radius of fireballs (especially for the
pulses in a burst), the shorter pulses have smaller value of p (Here
$p=\Gamma^{-2}\Delta\tau_{\theta,\rm FWHM}$), which might implicate
that short pulses come from larger $\Gamma$ and narrower local
pulses than long pulses, and short bursts come from largest $\Gamma$
and narrowest local pulses than long bursts, this is a reasonable
result if short GRBs are likely to be produced by the merger of
compact objects while long GRBs result from the core collapse of
massive stars. (3)All GRBs may arise intrinsically from local
Gaussian pulses, and these local pulses' widths,
$\Delta\tau_{\theta, FWHM}$, would not be less than 0.1, which is in
agreement with those found in Qin et al. (2005b). (4)The observed
pulses deviated down from the the dead line of local Gaussian pulse
would arise from the narrower local pulses (i.e., smaller than 1),
and the shorter the local pulses, the more obvious the deviation
feature. However we suspect that there will not be such a observed
pulse far deviated up from the dead line of the local Gaussian pulse
in the $t_r-t_{\rm FWHM}$ plane if it arises from a local Gaussian
pulse in terms of the fireball model.

These features above may provide constraints on intrinsic emission
mechanism responsible for GRBs.
\begin{acknowledgements}
This work was supported by the Special Funds for Major State Basic
Research Projects (``973'') and National Natural Science
Foundation of China (No. 10273019 and 10463001).
\end{acknowledgements}

\begin{table}[]
  \caption[]{ The Coefficients For Equation (3) And (4)  }
  \label{Table1}
  \begin{center}\begin{tabular}{clcl}
  \hline\noalign{\smallskip}
Local pulse forms &  $k_{1}$      & k                    \\
  \hline\noalign{\smallskip}
$\mu=1$ of equation(2) & 0.412 $\pm$ 0.003  &  0.827 $\pm$ 0.022\\
$\mu=2$ of equation(2) & 0.597 $\pm$ 0.006  &  1.335 $\pm$ 0.036\\
$\mu=3$ of equation(2) & 0.717 $\pm$ 0.009  &  1.718 $\pm$ 0.041\\
An exponential rise and decay pulse & 0.306 $\pm$ 0.001  &  2.006 $\pm$ 0.122\\
A Gaussian pulse & 0.650 $\pm$ 0.007  &  2.782 $\pm$ 0.149\\
A rectangle pulse & 0.149 $\pm$ 0.001  &  0.318 $\pm$ 0.013\\
  \noalign{\smallskip}\hline
  \end{tabular}\end{center}
\end{table}

\begin{figure}
   \vspace{2mm}
   \begin{center}
   \hspace{3mm}\psfig{figure=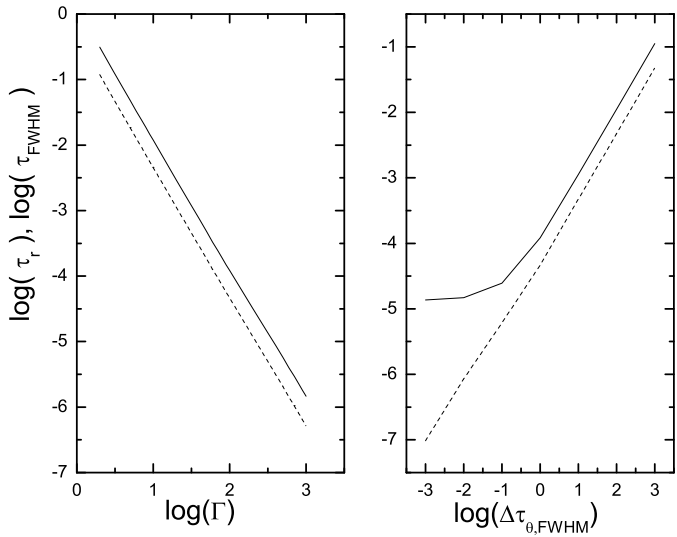} 
\caption{Relationships between $\tau_r$, $\tau_{\rm FWHM}$ and
$\Gamma$ (Left panel) and that between $\tau_r$, $\tau_{\rm FWHM}$
and $\Delta \tau _{\theta ,\rm FWHM}$ (Right panel) for the light
curves determined by equation (1), where a band function rest frame
radiation form with $\alpha _0=-1$ and $\beta _0=-2.25$, within the
frequency range of 100 $\leq$ $\nu/\nu_{0,p}$$ \leq$ 300, is
adopted, and we take $2\pi R_{\rm c}^3I_0/hcD^2=1$, $\mu=2$, $\tau
_{\theta ,\rm \min }=0$, $\theta_{\rm min}=0$, $\theta_{\rm
max}=\pi/2$, $\Delta \tau _{\theta ,\rm FWHM}$=1 (Left panel), and
$\Gamma$=100 (Right panel). The solid and the dash line represent
the $\tau_{\rm FWHM}$ and the $\tau_r$ of light curves in both
panels, respectively. }
   \label{Fig. 1}
   \end{center}
\end{figure}

\begin{figure}
   \vspace{2mm}
   \begin{center}
   \hspace{3mm}\psfig{figure=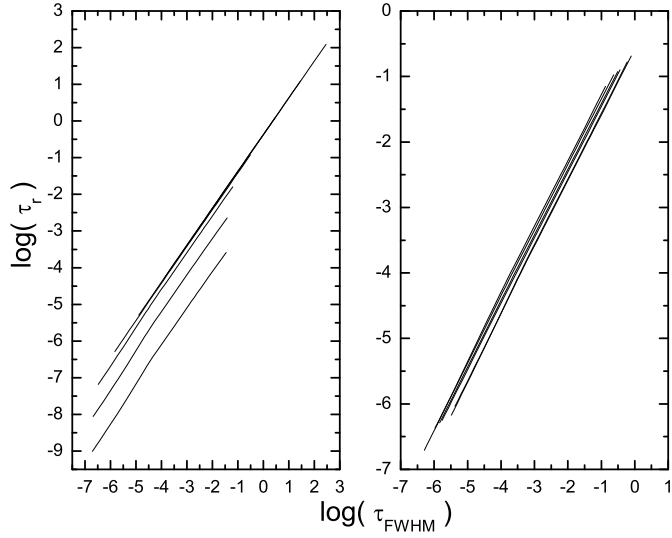} 
\caption{ Relationships between $\tau_r$ and $\tau_{\rm FWHM}$ for
the light curves determined by equation (1). Left panel: We take
$\Gamma$=2 to 1000 for different values of $\Delta \tau _{\theta
,\rm FWHM}$. The solid lines from the bottom to the top represent
$\Delta \tau _{\theta ,\rm FWHM}$=0.001, 0.01, 0.1, 1, 10, 100, and
1000, respectively. (note: when $\Delta \tau _{\theta ,\rm
FWHM}\geq1$, their corresponding lines overlap each other.) Right
panel: the dead lines of the six local pulse forms: the solid lines
from the bottom to the top represent local exponential rise and
exponential decay pulse, local Gaussian pulse, local power law pulse
$\mu$=3, 2, 1 of equation (2) and local rectangle pulse,
respectively. Other parameters are the same as those adopted in Fig.
1.  }
   \label{Fig. 2}
   \end{center}
\end{figure}

\begin{figure}
   \vspace{2mm}
   \begin{center}
   \hspace{3mm}\psfig{figure=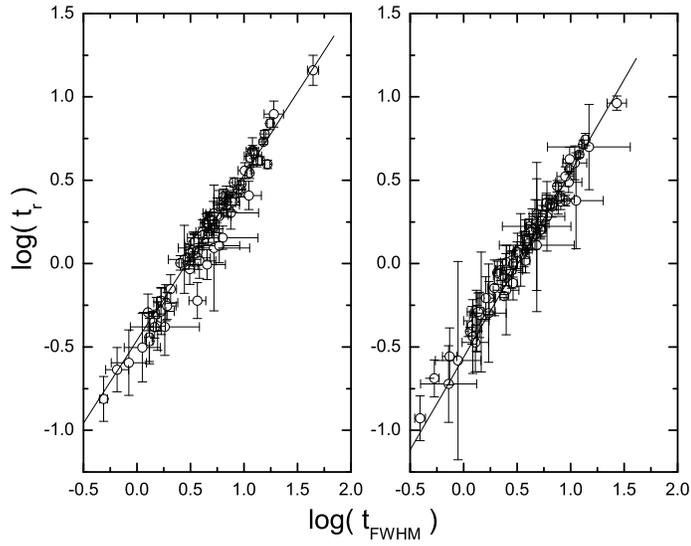} 
\caption{ Relationships between $t_r$ and $t_{\rm FWHM}$ for the
observed pulses based on sample 1. The left and the right panel
present the pulses of the 2nd channel and the 3rd channel,
respectively. The two solid lines are the fit lines of their data. }
   \label{Fig. 3}
   \end{center}
\end{figure}

\begin{figure}
   \vspace{2mm}
   \begin{center}
   \hspace{3mm}\psfig{figure=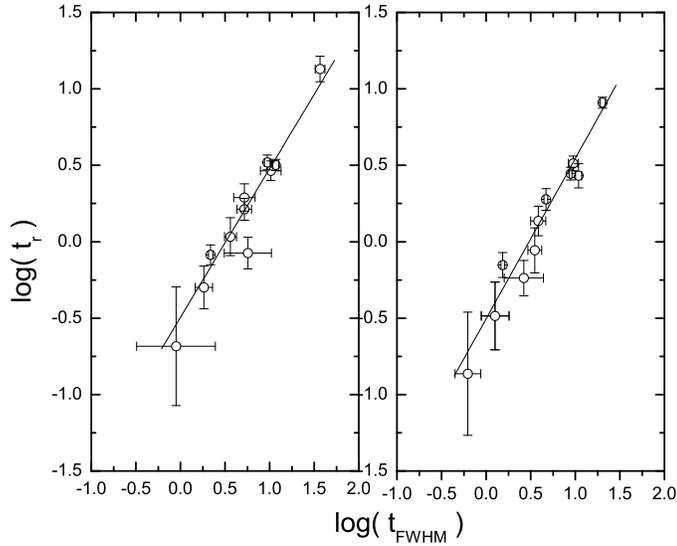} 
\caption{  Relationships between $t_r$ and $t_{\rm FWHM}$ for the
observed pulses based on sample 2. The left panel and the right
panel present the pulses of the band B (7-40 keV) and C (30-400
keV), respectively. The two solid lines are the fit lines of their
data. }
   \label{Fig. 4}
   \end{center}
\end{figure}

\begin{figure}
   \vspace{2mm}
   \begin{center}
   \hspace{3mm}\psfig{figure=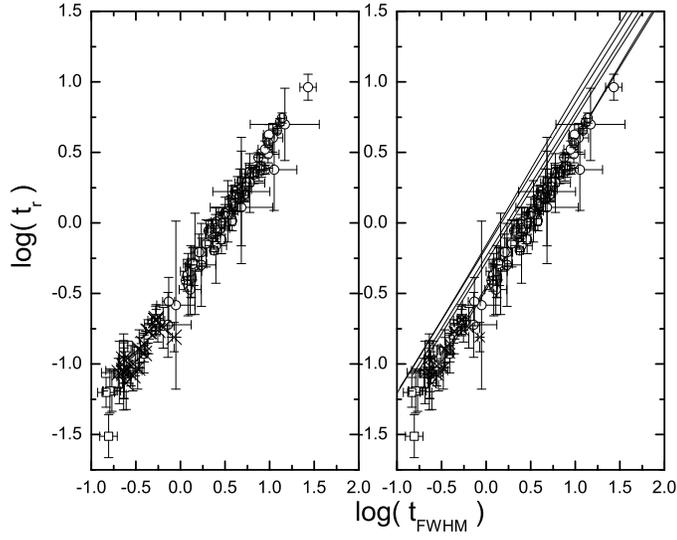} 
\caption{Relationships between $t_r$ and $t_{\rm FWHM}$ for the
observed pulses. Left panel presents the observed pulses of the 3rd
channel based on sample 1, 3 and 4. The open circle, the open
rectangle and  the cross present the pulses of sample 1, 3 and 4,
respectively. Right panel is a combination of the left panel and the
right panel in Fig. 2, where we take $R_{\rm c}=3\times10^{15}$cm. }
   \label{Fig. 5}
   \end{center}
\end{figure}
%
%
\label{lastpage}

\end{document}